\documentclass[preprint,12pt]{elsarticle}
\usepackage{amssymb}
\usepackage{amsmath}
\usepackage{listings}
\usepackage[ruled, vlined]{algorithm2e}
\usepackage{color}
\usepackage[english]{babel}

\newcommand*{\vv}[1]{\vec{\mkern0mu#1}}
\definecolor{grey}{rgb}{0.9,0.9,0.9}

\newcounter{bla}

\journal{Computer Physics Communication}

\begin{document}
\begin{frontmatter}
\title{RGIsearch: A C++ program for the determination of Renormalization Group Invariants}

\author{Rob Verheyen\corref{cor1}\fnref{label2}} 
\ead{RobVerheyen@gmail.com}
\cortext[cor1]{Corresponding author}
\fntext[label2]{IMAPP, Mailbox 79, Radboud University Nijmegen, P.O. Box 9010, 6500 GL Nijmegen, The Netherlands}

\begin{abstract}
RGIsearch is a C++ program that searches for invariants of a user-defined set of renormalization group equations. 
Based on the general shape of the $\beta$-functions of quantum field theories, RGIsearch searches for several 
types of invariants that require different methods. Additionally, it supports the computation of invariants up to two-loop level. 
A manual for the program is given, including the settings and set-up of the program, as well as a test case.
\end{abstract}

\end{frontmatter}

\newpage

{\bf Program Summary}
\begin{small}
\noindent
{\em Manuscript Title:}   RGIsearch: A C++ program for the determination of Renormalization Group Invariants\\
{\em Authors:}  Rob Verheyen\\
{\em Program Title:} RGIsearch\\
{\em Journal Reference:}\\
{\em Catalogue identifier:}\\
{\em Licensing provisions:} none\\
{\em Programming language:} C++\\
{\em Computer:} Desktop PC\\
{\em Operating system:} Linux\\
{\em RAM:} Ranging from several MB to several GB
{\em Classification:} 2.9 Theoretical Methods, 4.8 Linear Equations and Matrices, 5 Computer Algebra\\
{\em Nature of problem:} The determination of renormalization group invariants, which can be used as a probe for the high-energy behaviour of theories in particle physics.\\
   \\
{\em Solution method:} Several types of invariants are considered based on the general shape of the renormalization group equations. The problem of computing these invariants 
is then reduced to creating and solving very large systems of linear equations or generalized eigenvalue problems. Using the specific sparsity structure of these systems, the systems
can be solved.\\
   \\
{\em Restrictions:} Since the algorithms that solve the linear systems and generalized eigenvalue problems are very efficient, the main restriction is the amount of available RAM, where the systems are stored. The program supports polynomial renormalization group equations, which are typical particle physics, up to two-loop order.\\
   \\
{\em Running time:} Ranging from seconds to minutes.\\
   \\

\end{small}

\newpage

\section{Introduction}
Modern theories in particle physics, such as supersymmetry, often predict new physics at experimentally inaccessible energy scales. The high energy
behavior of such a theory can be probed by considering the evolution of its parameters, which were measured at some experimental scale, 
to higher energies using the \emph{renormalization group} (RG) equations.
Low-energy parameters can be translated to high-energy ones with several methods, each with their
own up- and downsides \cite{Hetzel:2012bk}. One of these methods uses RG \emph{invariants}, combinations
of the parameters of a theory that are invariant under the renormalization
group flow \cite{Demir:2004aq,Carena:2010gr}. While this method offers multiple advantages over others, finding
such invariants is highly nontrivial, especially at high loop orders or for theories with
large parameter spaces. RGIsearch is a C++
program that is able to compute these invariants for any set of RG equations. In this article, the methods that RGIsearch uses to find invariants are described and a manual is provided. Obviously, checking if the proposed invariants produced by the program are actually invariant is trivial. 
RGIsearch was recently used to compute several new invariants
of various versions of the Minimal Supersymmetric Standard Mode \cite{MSSM}.

\section{Methodology}

The RG equations are differential equations for the parameters of
a theory as function of the energy scale $\mu$. Consider the RG equation
of a renormalized theory for a parameter $x(\mu)$. The corresponding
$\beta$-function is defined as:

\begin{equation}
\beta(x)=16\pi^{2}\frac{dx}{dt},
\end{equation}

where $t=\mbox{log}_{10}\left(\mu/\mu_{0}\right)$. Since the $\beta$-functions
of quantum field theories are polynomials in the parameters of the
theory, RGIsearch considers several types of polynomial RG invariants.
We first discuss methods for these types for one-loop $\beta$-functions,
where in quantum field theories the coefficients of these polynomials are rational numbers, before extending
it to two loops.

\subsection{Monomial Invariants}

Monomial invariants assume the form:

\begin{equation}
M=\prod_{i=1}^{n}x_{i}^{a_{i}},
\end{equation}

where $x_{i}$ are the $n$ parameters of the theory and $\vec{a}\in\mathbb{Z}^{n}$. The powers $\vec{a}$ can
be taken to be integers since the coefficients of the $\beta$-functions
are rational, and any power of a monomial invariant is still a monomial
invariant. Taking the derivative and dividing out $M$, the condition for invariance
is:

\begin{equation} \label{monoCond} 
\sum_{i=1}^{n}\frac{a_{i}\beta(x_{i})}{x_{i}}=0.
\end{equation}

The left-hand side of eq. \eqref{monoCond} is a polynomial which is required to
vanish for all values of all $x_{i}$. This is only possible if all
monomial terms are cancelled internally. Therefore, eq. \eqref{monoCond} can be
reduced to a relatively small linear system of equations on the powers
$a_{i}$, where the equations represent the requirement of the cancellation
of all monomial terms. Every element in the nullspace of this linear
system represents a monomial invariant. By finding the basis vectors
for the nullspace, the independent monomial invariants are determined.
The other elements of the nullspace are then just products of powers
of these invariants.

\subsection{Dimensionalities}

Before extending the above method to polynomial invariants, the concept
of \emph{dimensionalities }has to be introduced. A dimensionality
$D$ is represented by a vector $\vv{D}\in\mathbb{Z}^{n}$ such that:

\begin{equation}
\mbox{dim}_{D}(x_{i})=D_{i}.
\end{equation}

Dimensionalities are additive for monomials:

\begin{equation}
\mbox{dim}_{D}\left(\prod_{i=1}^{n}x_{i}^{a_{i}}\right)=\vec{a}\cdot\vv{D},
\end{equation}

where $\cdot$ represents the standard Euclidean inner product. A
system of $\beta$-functions is said to have dimensionality $D$ if:

\begin{equation} \label{dims}
\forall i\in\left[1,...,n\right]:\:\mbox{dim}_{D}(\beta(x_{i}))-\mbox{dim}_{D}(x_{i})=c_{D},
\end{equation}

where the dimensionality of a polynomial such as the $\beta$-functions
means that all included monomial terms have the same
dimensionality, and $c_{D} \in \mathbb{Z}$ is a constant. A system of $\beta$-functions can have multiple dimensionalities. Note that these dimensionalities are similar to the well-known physical dimensionalities, such as mass dimension. In fact, mass dimension is typically one of the dimensionalities of a system of $\beta$-functions with $c_{D_{m}}=0$.
RGIsearch finds the dimensionalities of a system of $\beta$-functions
by reducing the problem to a different system of linear equations. Writing
out the $\beta$-functions in their monomial terms:

\begin{equation}
\beta(x_{i})=\sum_{j=1}^{m_{i}}C_{ij}M_{ij}\mbox{ where }M_{ij}=\prod_{k=1}^{n}x_{k}^{b_{ij,k}},
\end{equation}

where the $m_{i}$ are the number of monomial terms in $\beta(x_{i})$. Eq. \eqref{dims} can now be
written as:

\begin{equation}
\forall i\in\left[1,...,n\right],\,j\in\left[1,...,m_{i}\right]:\:\vec{b}_{ij}\cdot\vv{d}-D_{i}-c_{D}=0,
\end{equation}

which is a linear system in $\vv{d}$ and $c_{D}$ and can thus easily
be solved.

\subsection{Polynomial Invariants}

Similar to the $\beta$-functions, polynomial invariants assume the
form:

\begin{equation}
P=\sum_{i=1}^{m}C_{i}M_{i}\mbox{ where }M_{i}=\prod_{j=1}^{n}x_{j}^{a_{i,j}},
\end{equation}

where $\vv{C}\in Z^{m}$ and $\vec{a}_{i}\in Z^{n}$. Eq. \eqref{dims} implies:

\begin{equation}
\mbox{dim}_{D}\left(\frac{d}{dt}M_{i}\right)=\mbox{dim}_{D}\left(M_{i}\right)+c_{D}.
\end{equation}

This means that there cannot be any overlap between derivatives of
monomial terms with different dimensionalities. Therefore, it suffices
to consider polynomial invariants with monomial terms of the same
dimensionalities. Thus, RGIsearch first computes the dimensionalities
of the system. Next, if $r$ dimensionalities are found, it runs through
a user-defined range of dimensionalities $\vv{d}\in\mathbb{Z}^{l}$,
computing the set:

\begin{equation}
\mathcal{M}_{p}(\vv{d})=\left\{ \prod_{i=1}^{n}x_{i}^{a_{i}}|\forall l\epsilon\left[1,...,r\right]:\;\mbox{dim}_{l}\left(\prod_{i=1}^{n}x_{i}^{a_{i}}\right)=d_{l},\:-p\leq x_{i}\leq p\right\} ,
\end{equation}

where $p$ is a user-defined setting controlling the size of $\mathcal{M}_{p}(\vv{d})$.
The polynomial invariant then becomes:

\begin{equation}
P(\vv{d})=\sum_{i=1}^{|\mathcal{M}_{p}(\vv{d})|}C_{i}M_{i}\mbox{ where }M_{i}\in\mathcal{M}_{p}(\vv{d}).
\end{equation}

The requirement for invariance now is:

\begin{equation} \label{poly}
\sum_{i=1}^{m}C_{i}\frac{d}{dt}M_{i}=0,
\end{equation}

which can be converted to a linear system of equations in $\vv{C}$ by demanding that eq. \eqref{poly} holds for all values of all parameters.
While the systems encountered previously were relatively small, the
size of this system heavily depends on the size of $\mathcal{M}_{p}$,
which should be taken as large as possible to reach more invariants.
The associated system of equations can grow very large, but it can still be solved quickly due to its sparsity. 

The elements of the nullspace make up all possible invariants that
can be constructed from the set $\mathcal{M}_{p}$. By computing the
basis vectors of the nullspace, RGIsearch finds all independent invariants.
The rest of the nullspace consists of linear combinations of these
invariants. In addition, products of these invariants can be found
at other dimensionalities. After running through the range of user-defined
dimensionalities, a simple filtering algorithm searches for these
products of invariants and removes them from the output.

\subsection{Factorization}

RGIsearch supports a minor extension to the method of the previous
section, where a single variable can be factorized from a polynomial
invariant:

\begin{equation}
P_{j}\left(\vv{d}\right)=x_{j}^{b}\left(\sum_{i=1}^{m}C_{i}M_{i}\right)\mbox{ where }M_{i}\in\mathcal{M}_{p}(\vv{d}).
\end{equation}

Taking the derivative and dividing out $x_{j}^{b}$, the requirement
for invariance is:

\begin{equation} \label{factorize}
\frac{b\beta(x_{j})}{x_{j}}\sum_{i=1}^{m}C_{i}M_{i}+\sum_{i=1}^{m}C_{i}\frac{d}{dt}M_{i}=0.
\end{equation}

Instead of a linear system, eq. \eqref{factorize} leads to a system of the form:

\begin{equation}
 \mathbf{A} +  b\mathbf{Q} = 0.
\end{equation}

This can be interpreted as a nonsquare, generalized eigenvalue problem of a similar size as regular polynomial searching. It is approached as a regular eigenvalue system by solving the eigenvalues $b$ first, after which the system becomes linear. 

\subsection{System Creation}

The methods described in subsections 2.3 and 2.4 require the creation of very large  (typically $\mathcal{O}(10^{5})\times\mathcal{O}(10^{4})$
to $\mathcal{O}(10^{6})\times\mathcal{O}(10^{5})$) matrices. 
These matrices are constructed by grouping together monomial terms that appear in eq. \eqref{poly} and eq. \eqref{factorize}. 
RGIsearch performs this task by calculating these derivative monomial terms one-by-one and directly placing them into the matrix. 
To be able to do that, a \lstinline[language=c++, basicstyle=\ttfamily]{std::map} is used to make the translation from a monomial term to a row in the matrix. 

Monomial terms are stored in a \lstinline[language=c++, basicstyle=\ttfamily]{std::vector} of small integer numbers indicating the parameters and their powers. 
Because multiple numbers are required to fully specify a monomial term, and because containers such as \lstinline[language=c++, basicstyle=\ttfamily]{vector} 
require some memory overhead\footnote{Containers without overhead exist, but their basic functions such as \lstinline[language=c++, basicstyle=\ttfamily]{size()}, which is heavily used throughout the algorithm, have longer computation times, slowing down the algorithm significantly.}, the \emph{Cantor pairing function}:

\begin{equation}
P:\mathbb{N}\times\mathbb{N}\rightarrow\mathbb{N}\mbox{ where }P\left(n_{1},n_{2}\right)=\frac{1}{2}\left(n_{1}+n_{2}\right)\left(n_{1}+n_{2}+1\right)+n_{2},
\end{equation}

is used to convert monomial terms into a single natural number, which is then mapped to a row in the matrix. To pair more than two numbers, the \lstinline[language=c++, basicstyle=\ttfamily]{vector} is extended with zeroes until its size is a power of two. The pairing function is then applied pairwise until a single number remains. The Cantor pairing function is injective, so in principle  there is no danger of collision. However, the resulting number might overflow its designated 128 bit integer. Even if this happens, the probability of a collision is negligibly small.\footnote{The calculation of this probability is comparable to the birthday problem. The probability of collision for a 128 bit integer and a matrix of typical size can be approximated to $1-\mbox{exp}(-10^{12}/2^{129})$} 

\subsection{Markowitz Pivoting and Sparse Matrix Handling}

Next, RGIsearch solves the systems that were created in the above procedure. 
To find the nullspace of these systems, RGIsearch uses fraction free
Gaussian elimination with Markowitz pivoting. An implicit pivoting step
is performed before every elimination step to preserve sparsity. This
pivoting is based on the heuristic \emph{Markowitz count} \cite{Duff:1986:DMS:18753}:

\begin{equation}
\left(r_{i}-1\right)\left(c_{j}-1\right),
\end{equation}

where $r_{i}$ is the number of nonzero elements in row $i$ and $c_{j}$
the number of nonzero elements in column $j$. Before every elimination
step, the nonzero element with the lowest Markowitz count is located
and (implicitly) pivoted to the top-left. By performing this pivoting step, 
sparsity is preserved and and the number of required operations is reduced as much as possible. 

Note that the Markowitz count is zero for rows or columns with only a single nonzero element
(singlets). These rows and columns are selected first, since they
are very cheap to eliminate and produce no additional fill-in. The matrices that appear while computing
polynomial invariants usually contain many row singlets, so the algorithm
is specifically tailored to use these. It is abstractly
shown below.. The $i$th row of a matrix $\mathbf{A}$ is denoted with
$A_{i}$, and the matrix is viewed as a collection of these rows since
the order of the rows can freely be changed in nullspace computations. 

\begin{algorithm}[]
\DontPrintSemicolon

\renewcommand{\thealgocf}{}
\caption{Gaussian elimination with Markowitz pivoting}
\SetKwFor{ForAll}{for all}{do}{endfch}
\SetKwInOut{Input}{Input}\SetKwInOut{Output}{Output}
\SetArgSty{textnormal} 

\Input{A sparse matrix $\mathbf{A}\in\mathbb{\mathbb{Z}}^{m\times n}$}
\Output{A fully eliminated matrix $\mathbf{B}\in\mathbb{Z}^{r\times n}$,
a vector of flags that indicate elimination of variables $\vv{o}\in\mathbb{Z}_{2}^{n}$ }

$B\leftarrow\emptyset$, $Q\leftarrow\emptyset$, $\vv{o}\leftarrow\left(0,...,0\right)$ \;
\ForAll{$k \in [0,...,m] $}
	{
	\If{$A_{k}$ is a row singlet}
		{
		$Q\leftarrow Q\cup\left\{ k\right\}$ \;
		}
	}
\;
\While{$\mathbf{A}$ is not empty or $\vec{o}$ still contains zeroes}
	{
	$i\leftarrow0$, $j\leftarrow0$ \;
	\eIf {$Q\neq\emptyset$}
		{
		$ i\leftarrow$ Some element of $Q$ \;
		$Q\leftarrow Q-\left\{ i\right\} $ \;
		\If{$A_{i}$ is a row singlet}
			{
			$j\leftarrow$ Column index of the nonzero element of $A_{i}$\;
			}
		}
		{
		Find the nonzero element $a_{mn}$ in $\mathbf{A}$ minimizing $\left(r_{m}-1\right)\left(c_{n}-1\right)$\;
		$i\leftarrow m$, $j\leftarrow n$\;
		}
	\;
	\If{$j\neq0$}
		{
		$o_{j}\leftarrow1$ \;
		$B\leftarrow B\cup\left\{ A_{i}\right\} $\;
		\ForAll {$l$ with $a_{lj}\neq0$}
			{
			Fraction-freely eliminate $a_{lj}$ from $A_{l}$ using row $A_{i}$ \;
			Reduce $A_{p}$ using the gcd of its elements \;
			\If {$A_{l}$ is empty}
				{
				$A\leftarrow A-\left\{ A_{l}\right\} $\;
				}
			\If {$A_{l}$ is a row singlet}
				{
				$Q\leftarrow Q\cup\left\{ l\right\} $
				}
			}
		}
	$A\leftarrow A-\left\{ A_{i}\right\} $ 
	}

\end{algorithm}

The reduction of a row simply means that all nonzero elements of the
row are divided by their total greatest common divisor (gcd). If the
algorithm terminates while $\mathbf{A}$ is not yet empty, then all
variables have been eliminated and no nullspace exists. If $\mathbf{A}$
does empty, the nullspace can easily be solved by selecting a basis
for the space of variables that have not been eliminated. The rest
of the variables are then backwardly solved by the equations in $\mathbf{B}$. 

For the method described in subsection 1.4, elimination of matrices
of similar sizes containing univariate polynomials with integer coefficients
has to be performed. The algorithm is very similar. The Markowitz
count is extended to:

\begin{equation}
\left(r_{i}-1\right)\left(c_{j}-1\right)\left(d_{ij}-1\right),
\end{equation}

where $d_{ij}$ is the degree of the matrix element at row $i$ and
column $j$. Furthermore, integer coefficients are maintained through
fraction-free Gaussian elimination using the subresultant polynomial 
remainder sequence algorithm outlined in \cite{Brown:1978:SPA}. 
The eigenvalues are then computed by solving the polynomials 
that appear on the implicit diagonal of $\mathbf{B}$. 

To improve speed and memory efficiency, RGIsearch uses the \emph{list
of lists }method to store matrices \cite{Duff:1986:DMS:18753}.
They are stored as a list of rows, containing all nonzero matrix
elements which are tagged by their column index. While this method
is not the most efficient of all sparse matrix storage techniques in terms
of memory usage, it allows for very fast access to individual rows which
is extremely important in the above algorithm. Additionally, insertion
of new nonzero matrix elements is relatively cheap which is significant
for the elimination steps. Fast access to columns is also important 
during the elimination steps. Therefore, a second matrix is stored 
as a list of columns instead of rows. This matrix is used purely as an 
indexing structure and thus only stores the row indices of nonzero elements. 
Both structures are updated simultaneously during the elimination procedure.

\subsection{Two-loop invariants}

RGIsearch allows for computation of regular polynomial invariants
for two-loop $\beta$-functions. The general form of a two-loop $\beta$-function
is:

\begin{equation}
\beta\left(x_{i}\right)=\beta^{(1)}\left(x_{i}\right)+\frac{1}{16\pi^{2}}\beta^{(2)}\left(x_{i}\right).
\end{equation}

Therefore, a two-loop invariant looks like:

\begin{equation}
I=I_{1}+\frac{1}{16\pi^{2}}I_{2}.
\end{equation}

Taking the derivative and letting $I_{1,2}^{(j)}$ denote the part
of the derivative involving $\beta^{(j)}(x_{i})$, the condition for
invariance for the one-loop and two-loop terms separate:

\begin{equation}
I_{1}^{(1)}=0\mbox{ and }I_{1}^{(2)}+I_{2}^{(1)}=0.
\end{equation}

The term $I_{2}^{(2)}$ is formally of three-loop order. These conditions are converted into a linear system in
a very similar fashion to the one-loop case.

\section{Usage}

RGIsearch is available from a public repository \cite{github}. The latest version of RGIsearch can be acquired with:
\begin{lstlisting}[language=bash, backgroundcolor=\color{grey}]
$ git clone https://github.com/rbvh/RGIsearch.git
\end{lstlisting}

\subsection{Compilation}

To compile, do:

\begin{lstlisting}[language=bash, backgroundcolor=\color{grey}]
$ make
\end{lstlisting}

This will generate all binary files and link them into an executable. This executable
can then be run with:

\begin{lstlisting}[language=bash, backgroundcolor=\color{grey}]
$ ./RGIsearch
\end{lstlisting}

If any $\beta$-functions, settings or any of the sources are changed,
\colorbox{grey}{\lstinline{make}} 
should update the binaries. If for some reason it does
not, or it is otherwise required, one can do:

\begin{lstlisting}[language=bash, backgroundcolor=\color{grey}]
$ make clean
$ make
\end{lstlisting}

\subsection{Settings}
The program settings can be found in the file \colorbox{grey}{\lstinline{settings.h}}. 

\begin{lstlisting}[language=c++, basicstyle=\ttfamily]
const int DIM_SEARCH_PARAMETER
\end{lstlisting}

This constant controls the range of dimensionalities the algorithm searches. Depending on the dimensionalities that are found and the value of this constant, RGIsearch calculates a suitable selection of dimensionality vectors to search through. 

\begin{lstlisting}[language=c++, basicstyle=\ttfamily]
const int MAX_TERM
\end{lstlisting}

This constant controls the number of different parameters allowed in a monomial term during the execution of the algorithms for polynomial invariants. For instance, if it is set to 2, terms like $xy^2$ can occur, but $xyz$ is forbidden. Changing it can have huge influence of the required computation times and memory.

\begin{lstlisting}[language=c++, basicstyle=\ttfamily]
const int FILTER_THRESHOLD
\end{lstlisting}

This constant controls the filtering mechanism. If it is set to a higher value, the filtering algorithm will perform a more elaborate attempt to find all possible products of previously found invariants to compare against any newly found invariants. The default value should almost always be sufficient.

\begin{lstlisting}[language=c++, basicstyle=\ttfamily]
bool INCLUDE_TWO_LOOP
\end{lstlisting}

Set to true to search for invariants at two-loop level. 

\begin{lstlisting}[language=c++, basicstyle=\ttfamily]
bool FACTORIZE
\end{lstlisting}
Set to true to include the algorithm that searches for factorized polynomial invariants. The code does currently not support factorization for two-loop level searches.

\begin{lstlisting}[language=c++, basicstyle=\ttfamily]
bool REPORT
\end{lstlisting}

Set to true to make the program report its activities in more detail.

\subsection{$\beta$-Functions}
The $\beta$-functions can be found in \colorbox{grey}{\lstinline{equations.cpp}}. 

To define the $\beta$-functions of a theory, the parameters of the theory first need to be defined. This is done with:

\begin{lstlisting}[language=c++,basicstyle=\ttfamily]
Param newPar("parName", parSize);
\end{lstlisting}

\lstinline[language=c++, basicstyle=\ttfamily]{parSize} is the size of the parameter ($1$ for a scalar, $n$ for a $n\times n$ matrix).
If a parameter is complex, its daggered counterpart must be defined. This is done with:

\begin{lstlisting}[language=c++,basicstyle=\ttfamily]
Param newParDagger = dagger(newPar);
\end{lstlisting}

The \lstinline[language=c++, basicstyle=\ttfamily]{ Param} class has several methods that simplify matrix parameters. They include:
\begin{enumerate}
\item \lstinline[language=c++, basicstyle=\ttfamily]{botRight()} sets all elements except for the bottom right component equal to zero.
\item \lstinline[language=c++, basicstyle=\ttfamily]{diag()} sets all offdiagonal elements equal to zero.
\item \lstinline[language=c++, basicstyle=\ttfamily]{botRightDiag()} sets all offdiagonal terms equal to zero, and makes all components except for the bottom right degenerate.
\end{enumerate}

After defining the parameters, one can define the $\beta$-functions. These are stored as a vector of the objects \lstinline[language=c++, basicstyle=\ttfamily]{BetaFunc}: 

\begin{lstlisting}[language=c++,basicstyle=\ttfamily]
vector<BetaFunc> nameOfBetaFuncs
\end{lstlisting}

The $\beta$-function of a parameter can then be defined as:
\begin{lstlisting}[language=c++,basicstyle=\ttfamily]
BetaFunc bNewPar(newPar);
bNewPar = <polynomial>;
nameOfBetaFuncs.push_back(bNewPar); 
\end{lstlisting}

The polynomial can be constructed using regular arithmetic involving the parameters of the theory. Irrational numbers $a/b$ can be represented as \lstinline[language=c++, basicstyle=\ttfamily]{ir(a,b)}. 
A trace function is available as \lstinline[language=c++, basicstyle=\ttfamily]{Tr()}. For the complex conjugate parameters, their $\beta$-functions must be included as:

\begin{lstlisting}[language=c++,basicstyle=\ttfamily]
BetaFunc bNewParDagger = Conjugate(bNewPar);
nameOfBetaFuncs.push_back(bNewParDagger);
\end{lstlisting}

RGIsearch needs separate \lstinline[language=c++, basicstyle=\ttfamily]{vector}s of $\beta$-functions for one-loop and two-loop, which must be in the same order. Finally, the algorithm can be initiated by calling:

\begin{lstlisting}[language=c++,basicstyle=\ttfamily]
findInvariants(1loopBetaFuncs, 2loopBetaFuncs);
\end{lstlisting}

\subsection{Test Program}

As an example, we consider a simple toy system of $\beta$-functions for three variables $x, y, z$:

\begin{subequations}
\begin{align}
\beta(x) &= -4xy -3y^{2} + \frac{1}{16\pi^{2}}(-x + 6y),\\
\beta(y) &= -2x^{2} + z + \frac{1}{16\pi^{2}}3x,\\
\beta(z) &= 6xy^{2} + 4yz + \frac{1}{16\pi^{2}}(4xy - 3y^{2} + z)
\end{align}
\end{subequations}

This system has two invariants at two-loop level. To find them, the appropriate \colorbox{grey}{\lstinline{equations.cpp}} is shown below. 

\begin{lstlisting}[language=c++, basicstyle=\ttfamily, numbers=left, numberstyle=\tiny, frame=tb, columns=fullflexible, showstringspaces=false, title=Test case \colorbox{grey}{\lstinline{equations.cpp}} file, float]
#include "src/common.h"
#include "src/invariants.h"

int main()
{
Param x("x", 1);
Param y("y", 1);
Param z("z", 1);

vector<BetaFunc> test_1;
vector<BetaFunc> test_2;

BetaFunc bx_1(x);
BetaFunc bx_2(x);
bx_1 = -4*x*y - 3*y*y;
bx_2 = -1*x + 6*y;
test_1.push_back(bx_1);
test_2.push_back(bx_2);

BetaFunc by_1(y);
BetaFunc by_2(y);
by_1 = -2*x*x + 1*z;
by_2 = 3*x;
test_1.push_back(by_1);
test_2.push_back(by_2);

BetaFunc bz_1(z);
BetaFunc bz_2(z);
bz_1 = 6*x*y*y + 4*y*z;
bz_2 = -4*x*y - 3*y*y + 1*z;
test_1.push_back(bz_1);
test_2.push_back(bz_2);

findInvariants(test_1, test_2);
}
\end{lstlisting}

Using the default settings and enabling two-loop calculations, the output is:

\begin{lstlisting}[language=bash, backgroundcolor=\color{grey}]
-x^2 + 2y^2 - z + (1/16pi^2){x + y}
-2y^3 - 2xz + (1/16pi^2){x^2 + 3z}
\end{lstlisting}

These can easily be verified to be invariants of the above system.

\section*{Acknowledgements}
Thanks to Ronald Kleiss and Wim Beekakker for suggestions and guidance along the way, and more so to Ronald Kleiss for carefully proofreading of the draft of this paper.
\clearpage
\bibliographystyle{elsarticle-num}
\bibliography{literature}

\end{document}